\documentclass[twocolumn,showpacs,preprintnumbers,amsmath,amssymb]{revtex4}

\usepackage{graphicx}
\usepackage{dcolumn}
\usepackage{bm}

\begin{document}

\preprint{APS/123-QED}

\title{The effect of electric fields on lipid membranes}

\author{Zlatko Vasilkoski}
 \altaffiliation[Email: ]{zlatko@neu.edu }
\affiliation{%
022 Dana Research Center, Northeastern University, Boston, MA 02115
}%


\date{\today}

\begin{abstract}
Contrary to existing theoretical models, experimental evidence points out that electroporation (membrane defect formation under external electric fields) starts to occur within the range of transmembrane voltages that cells may routinely experience, curiously, just above the range of transmembrane voltages involved in neural signal transmission. Understanding the underlying principles of electric fields-lipid membrane interactions seems to carry a great biological importance.

An argument is presented toward understanding the theoretical aspects of electroporation by using the DLVO theory, which has not been recognized previously in the context of electroporation. Further, the dispersion interactions (with its quantum nature), of the double layer counterions and membrane lipid molecules over the Stern layer are emphasized. The sign of these forces is such that they compress the membrane. A parallel is drawn to the theory of thin films. The argument is that the external electric field breaks the symmetry of the disjoining pressures on both sides of a lipid membrane, resulting in a protrusion of only few lipid molecules. That compromises the membrane stability on a nanoscale and makes it traversable to ions. The presented estimate based on these arguments is consistent to electroporation experiments and existing numerical simulations.

\end{abstract}

\maketitle

\section{\label{sec:level1}Introduction}

Foam films and membranes have been studied for over 100 years. They are a bilayer formation with or without an aqueous core respectively, in which the hydrophilic groups are facing the aqueous solution. In the case of membranes the hydrophobic ends are in opposite direction facing each other, and in the case of foam films, soap bubbles, for example, the hydrophobic ends are exposed to air. In both, foam films (soap lamellas) and double layer membranes (lipid membranes), an equilibrium on the order of molecular dimension thickness (nm) can be reached, creating black films and membranes \cite{Exerowa_1998}. The biomembranes play a vital role in the activities of organelles and cells, and they are very closely related to this group of bilayer formations. Their main function varies, but one of the essential roles is to provide an excellent barrier, allow a very complex and selective molecular signaling and transport, including propagation of neural signals. 

Beside diverse composition and thickness of the bilayer and its exposition to air or aqueous medium, they all exhibit more or less elastic stability. Under mechanical or electrical stress they rupture (lyse) and in some occasions they may exhibit reversibility of the defects (reversible breakdown). The reversible breakdown has been observed in the case of soap films exposed to electric fields for example \cite{Winnie_1969} and is a more common feature of lipid membrane electroporation. This indicates an underlying mechanism how these disperse systems can be analyzed and described. \\

\subsection{\label{sec:level2}Electroporation}

The electrical breakdown and reversible breakdown of biological membranes has been subject of intensive experimental and theoretical studies. However, the effects the electric fields have on membranes, and the actual mechanism of electrically induced instabilities is still not satisfactory understood. Some models are concerned with more macroscopic reasons for instability as lipid membrane thinning, related to electrostriction , undulation and elastic properties of the membranes \cite{Winterhalter_1987, Winterhalter_1988, Winterhalter_1998, Kozlov_1992}. Other models, as electroporation models, find the explanation in formation and expansion of tiny pores. This process of electrically creating aqueous pores in the membrane, large enough to be traversable to ions and molecules is termed electroporation. The effect is believed to be intrinsic to artificial planar bilayer membranes, as well as to any other biological membrane of a cell or an organelle. Thus, the electroporation process may strongly increases the ionic and molecular transport across a biologically relevant membranes, redistribute membrane's lipid content and cause many biologically significant effects. 

The process of electroporation has been routinely used in cell biology to load cells with small molecules (drugs), and large molecules (as proteins and DNA) \cite{WeaverChizmadzhevChapterCRC1996}.

Experimental evidence over many decades 
\cite{ BenzBeckersZimmermannChargePulsePlanarBilayersJMembraneBio1979} suggested that the electroporation process is a stochastic process, with pore size evolution and occurrence of reversible or irreversible breakdown \cite{ AbidorEtAlChizmadzhevMainFacts1stPaper1979, BarnettWeaverEporeUnifiedTheoryBreakdownRuptureBB1991}. It is believed that the pores completely dominate electrical conductance, but they occupy very small fractional area. 

Bilipid membrane thickness is much smaller than the wavelength of light so the process of electroporation is most commonly observed through change of electrical conductance of the membrane (believed as a result of electrically induced pores). Recent accurate measurements \cite{Melikov_2001} have recorded a change in conductance due to a single defect (pore) at transmembrane voltages as little as $\approx 250$ mV. The pore size estimates based on the change in conductance indicate a nanometer size defects, and a lifetime in the range of milliseconds.

The theoretical models of electroporation are roughly concerned with two distinct aspects, the theory of pore creation and the theory of pore evolution. 

The pore evolution concepts have roots \cite{ Derjaguin_1989, Derjaguin_1987,Evans_2003} in the work of J.B. Zeldovich, from 1943. Based on experiments and theoretical analysis, the Chizmadzhev group in 1979, has described the initial electroporation theory for artificial planar bilayer membranes in a set of seven papers \cite{AbidorEtAlChizmadzhevMainFacts1stPaper1979}. The use of Smoluchowski equation in the context of pore size evolution was given by \cite{AbidorEtAlChizmadzhevMainFacts1stPaper1979, PowellWeaverEporeTheoryBB1986}.

While the pore evolution and expansion models seem to be well established, the initial steps of pore formation are less certain. A central part of almost all electroporation models is the critical permealization threshold needed for pore creation. A phenomenological, Arrhenius like model of transmembrane voltage dependent energy barrier was independently proposed by \cite{ WeaverMintzerEporeTheoryBilayerStabilityPhysLett1981, SugarPoreLipidBilayerTheoryJPhysiolParis1981}. More recent electroporation simulations \cite{JoshiSchoenbach_PhysRev2000, DeBruinKrassowska_1998} use different pore evolution concepts, but a common Arrhenius energy barrier of $\approx 45-50$ kT for pore creation \cite{Vasilkoski_2006}, which is in agreement with the experimental values of critical voltages causing electroporation. The existing theoretical models, related to membrane elasticity, seem to predict critical voltages of several Volts causing membrane instability. That exceeds the experimental lower threshold values ($\approx 250$ mV - $1$ V) by far. Model modifications have been proposed \cite{Jordan_1998} to avoid this discrepancy, but it seems that mechanism of membrane's electrical breakdown is still not satisfactory understood.

Besides all this advances, it seems that many questions related to the physical mechanisms of the electroporation are still not well understood \cite{Teissie_2005}, and more basic research is needed. 

Based on the accumulated experimental observation and the recent measurements, specifically, \cite{Melikov_2001}, and molecular dynamics simulations \cite{Tieleman_2004}, a few main electroporation aspects, relevant for this paper, can be summarized.\\

- Relatively low transmembrane voltages of $\approx 200$ mV can cause electroporation. \\
- Poration is a local effect (just few lipids involved).\\
- Pores are nanometer size defects.\\
- It takes a short time (ns) to create a pore. \\
- It takes a long time (ms) to destroy a pore. \\
- The process is sensitive to concentration and pH. \\
- The process causes electroporation asymmetry and lipid redistribution on the cell membrane (flip-flop). \\

This paper, further mainly addresses the theoretical aspects of pore creation. The other aspects of the electroporation will be addressed in a future work. 


To put all this into context, a few basic aspects of lipid membranes and mean field theories of ion interactions with membranes, are briefly noted.

\subsection {\label{sec:level2} General about lipid membranes}
\subsubsection {\label{sec:level2} Lipid membrane fluctuations}

Based on spectroscopic data a time scale for reorientation of chemical bonds, molecular segments, molecules, or assemblies of molecules can be placed. In general, the correlation time increases with increasing size of moving segments. The time scale of lipid motion in biomembranes and their approximate correlation times cover almost $15$ orders of magnitude \cite{Yeagle_2005}. Protrusion, the head groups sticking in or out of the membrane plane is on the order of $10^{-9} $ s. Rotational diffusion of the head group or wobble is on a scale of $10^{-8}$ s. The lateral diffusion (in the plane of the membrane) is on the order of $10^-7$ s, with a diffusion constant of $D \approx 10^{-12}$ to $10^{-11}$ $\mathrm{m}^2/\mathrm{s}$, and lipids spanning distances of $\approx 10 \mu$m in $1$ s. The large scale collective deformations (curvatures) from the plane of the membrane - undulations are on a time scale of $10^{-6}$ s to $1$ s. They are thermally driven, and membrane's waves like motions correspond to very large wavelengths. Finally the jumps of lipids from one side on the membrane to the other or the so called flip-fop is dependent on molecule's polarity, and is on the order of $10^3$ s to $10^4$ s. 

\subsubsection{\label{sec:level3}The pressure profile in lipid membranes}

Protrusion of the lipid molecules is closely related to the lateral pressure profile. Most of the biologically relevant lipid bilayer membranes are made of lipid molecules with two hydrocarbon chains \cite{Israelachvili}. The lateral pressure profile of these lipid bilayer membranes is symmetric with respect to the middle of the membrane. Going towards the middle of the bilayer \cite{Israelachvili}, there is initially headgroup repulsion, and then interfacial attraction. Coming towards the two hydrocarbon chains, repulsion is encountered again, since hydrophobic hydrocarbon tails dislike each other's presence. All of these pressures are summed up to zero. 

The value of these pressures is high. Various molecular dynamics simulations \cite{Sonne_2005, Carillo_2005, Gullingsrud_2004, Lindahl_2000} indicate values in the range of hundreds of atmospheres. A simulation estimate using lattice statistical thermodynamic calculations of lateral pressure profiles \cite{Cantor_1999} give a value of over $400$atm. A lower threshold value of $\approx 400$atm will be used throughout this paper as a approximate threshold of membrane instability. 

Molecular dynamics simulations of electroporation \cite{Tieleman_2004} also indicate high values of pressures. The forces of tens of pN on the area of a lipid headgroup ($\approx 0.6$ nm$^2$)\cite{Israelachvili}, result in fairly large pressures of few hundred atmospheres. 


\subsection{\label{sec:level2}Mean Field Theory of Ions near lipid membrane}

The surface charge density of lipid membranes mainly comes from the lipid head groups that are exposed to the aqueous solution. The electric field of this surface charge redistributes the ions in the electrolyte due to the local concentration ($n$) and potential ($\psi$) gradients that can be described according to the familiar equation 

\begin{eqnarray}
\frac{{\partial \rho }}{{\partial t}} = \frac{\partial }{{\partial x}}\left[ {
D\frac{{\partial \rho }}{{\partial x}} + \frac{{ze}}{{kT}}\rho \frac{{\partial \psi }}
{{\partial x}}} \right]
\label{eq:SE},
\end{eqnarray}

where planar symmetry is assumed, $x$ being perpendicular to the membrane, and the charge density $\rho = \sum\limits_i {\rho _i =} \sum\limits_i {n_i z_i e}$ is a sum over all the concentrations of ions $n_i$, of valence $z_i$. At electrochemical equilibrium we have 

\begin{eqnarray}
\rho \left( x \right) = \sum\limits_i {\rho _{i0} e^{ - \frac{{z_i e}}{{kT}}\psi\left( x \right)} } 
\label{eq:eqbl},
\end{eqnarray}

where $\rho _{i0}$ is the bulk charge density far away from the membrane 
($\psi\left( x \right) \approx 0$). Then the potential of this ion distribution at equilibrium is given by the Poisson-Boltzmann equation 

\begin{eqnarray}
\frac{{d^2 \psi }}{{dx^2 }} =  - \frac{e}{{\epsilon \epsilon _0 }}
\sum\limits_i {z_i n_{i0} e^{ - \frac{{z_i e}}{{kT}}\psi \left( x \right)} } 
\label{eq:PB}.
\end{eqnarray}

Rearranging equation (\ref{eq:PB}) it can be seen that the following expression 
is constant with $x$. 

\begin{eqnarray}
\frac{d}{{dx}}\left[ {\sum\limits_i {n_i } kT - \frac{{\epsilon \epsilon _0 }}{2}\left( {\frac{{d\psi }}{{dx}}} \right)^2 } \right] = \frac{d}{{dx}}\left[ P(x) \right] = 0
\label{eq:ConstRo}.
\end{eqnarray}

The expression in the brackets (pressure) shows that the entropy of mixing (the osmotic pressure) of the electrolyte is influenced by the electric field from the membrane surface charges. As a result, entropically unfavorable, counterions are more densely distributed against the membrane, but in such a way that the total pressure is constant in any transverse cross section from the membrane. The expression (\ref{eq:ConstRo}) is dependent on the ion concentration, and by applying an additional electric field, there will be ion redistribution. It should be mentioned \cite{Derjaguin_1987} that it is not necessary to resort to Poisson-Boltzmann equation and any assumptions leading to it, to obtain expression (\ref{eq:ConstRo}). In most general case we would have 

\begin{eqnarray}
\frac{d}{{dx}}\left[ {-\int\limits_0^\psi {\rho \left( \psi \right)} d\psi - \frac{{\epsilon \epsilon _0 }}{2}\left( {\frac{{d\psi }}{{dx}}} \right)^2 } \right] = \frac{d}{{dx}}\left[ P(x) \right] = 0
\label{eq:ConstP}.
\end{eqnarray}

\section{DLVO theory and Membranes}

An important contribution to the problem of stability of thin films was given by the works that started 60 to 70 years ago by Derjaguin, Landau, Verwey and Overbeek (DLVO theory). The behavior of the system of electrolyte column sandwiched between any two surfaces is a consequence of the interplay between repulsive and attractive (van der Waals) forces on the interface. The balance of these forces and the thickness of the film correspond to an interaction energy minimum. 

In 1939, Derjaguin originally formulated the concept of a disjoining pressure for thin liquid films and verified experimentally its existence \cite{Derjaguin_1987} \cite{Derjaguin_1989}. Changes in the interfacial region that generate the disjoining pressure in a thin liquid film originate from intermolecular forces. It is customary to separate the various contributions of the disjoining pressure into different components, e.g., 

\begin{eqnarray}
\Pi= \Pi_\mathrm{e} + \Pi_\mathrm{w} + \Pi_\mathrm{ster} + \Pi_\mathrm{adh}+ \ldots\ 
\label{eq:Ptot},
\end{eqnarray}
where the first two, the electrostatic ($\Pi_\mathrm{e}$) and the dispersion ($\Pi_\mathrm{w}$) component, usually refer to DLVO theory. The other components as steric, short range structural forces, and adhesion forces play important role at molecular distances. The key assumption of equation (\ref{eq:Ptot}) is that the various contributions to the disjoining pressure are additive. However, it is not always clear that this assumption is valid and it may lead, in some cases, to anomalous results. In general, thin film stability is influenced most importantly, by the electrostatic ($\Pi_\mathrm{e}$) and the dispersion ($\Pi_\mathrm{w}$) components \cite{Derjaguin_1987} \cite{Derjaguin_1989}. \\

Further the disjoining pressure components of a single lipid membrane in solution are analyzed, by considering the case when one of the films is moved far away.

\subsection{\label{sec:level2}The electrostatic component of disjoining pressure $\Pi_\mathrm{e}$}

DLVO describes the interaction pressure of thin films.  There the diffuse layers of the two charged surfaces overlap, and cause interaction pressure between them. The hydrostatic pressure at some distance $x$ from the membrane is given by the constant term in equation (\ref{eq:ConstP}). The difference of the hydrostatic pressures in the bulk (absence of the other surface) and in the interlayer is the electrostatic component of the disjoining pressure $\Pi_\mathrm{e}$ \cite{Derjaguin_1987} \cite{Derjaguin_1989}.

\begin{eqnarray}
\Pi_\mathrm{e}=P_{D}(x)-P_{\infty}(x)
\label{eq:Pela}.
\end{eqnarray}

Here $P_{D}(x)$ is the pressure at point $x$ in the interlayer when the two surfaces are $2D$ apart, and $P_{\infty}(x)$ is the pressure at the same point $x$ when the two surfaces are separated.  

If the overlapping diffuse layers are identical, there is a symmetry plane at $x=D$ where $\psi = \psi_\mathrm{m}$ and $E = 0$. Considering the anions and cations, in a $1:1$ electrolyte such as $NaCl$ or $KCl$, and a Gouy-Chapman $\rho(\psi)$, the disjoining pressure from equation (\ref{eq:Pela}) will have a cosine hyperbolic dependence on the midplane voltage $\psi_\mathrm{m}$,

\begin{eqnarray}
\Pi_\mathrm{e} \left( {\psi_\mathrm{m}} \right) = -\int\limits_0^{\psi_\mathrm{m}} {\rho \left( \psi \right)} d\psi = -2 n_0 kT\left[ {ch\left( {\frac{{ze}}{{kT}}\psi_\mathrm{m} } \right) - 1} \right]
\label{eq:Pel}.
\end{eqnarray}

For a single film (membrane), when the other film is far away, but $D$ is still finite, the fact that the mid plane potential $\psi_\mathrm{m}$ is very small ($\psi_\mathrm{m} \approx 0$) is used, and the following approximate expression (as derived by Verwey and Overbeek in 1948 \cite{ VerweyOverbeek_1948}) can be used.

\begin{eqnarray}
\Pi_\mathrm{e} \left( {\psi_\mathrm{m}} \right) \approx -n_0 \frac{{\left( {ze} \right)^2 }}{{kT}}\psi_\mathrm{m}^2 
\label{eq:Pel_apx}.
\end{eqnarray}

The pressure is directed towards the membrane. 

Further, in this paper, arguments are presented that the expression (\ref{eq:Pel_apx}) should be used to describe the interaction of external electric fields with a lipid membrane in electroporation models and theory. The argument for this follows the analogy that the presence of electrode's electric fields has a similar effect as another charged surface being present far away.

\subsection{\label{sec:level2}The dispersion component of disjoining pressure $\Pi_\mathrm{w}$}

The van der Waals dispersion forces are caused by polarization due to fluctuating electromagnetic fields. Besides being dependent on geometry, they are also frequency dependent, and they relate to the refractive index and dispersion.

It can be shown, for example \cite{Milonni, Milonni_Aspect_2000} that van der Waals interaction of two neutral polarizable particles can be attributed to correlations in the fluctuating vacuum electromagnetic field, that induces fluctuating dipole moments in the particles, and the correlation of these moments is the der Waals interaction. The van der Waals interactions can be derived from the Maxwell equations and the additional (quantum) assumption that each mode of the electromagnetic field has a zero-point energy $\frac{1}{2}\hbar\omega$. \\

Calculating the dispersion forces, even for simple geometry, is not an easy task and few aspects like nonadditivity, relativistic retardation and the sign need to be addressed. \\

In rarified media like a gas or a solute in a solution, non-pair wise interactions are small and additivity holds, but in condensed media, the dispersion forces are not additive, and in general, the force between two molecules depends on the presence of other molecules. These non-pair wise interaction energies may be attractive or repulsive, depending on the geometrical arrangement of the particles involved, and are important to be considered if the intermolecular spacing is small. The Lifshitz formula and theory \cite{Derjaguin_1987, Derjaguin_1989, Milonni} used here to calculate the Maxwell stress tensor is macroscopic, and avoids the problems associated with non-additivity of the dispersion forces. \\

When the intermolecular spacing is large, the relativistic retardation, due to the finite speed of light, needs to be included. This aspect plays an important role in colloidal systems and reduces the strength of van der Waals interaction \cite{Casimir_Polder_1948}. In the limiting case of perfect conductor, including retardation, the Lifshitz formula reduces to the Casimir force, which is a macroscopic manifestation of the van der Waals force. In either of these cases van der Waals interactions may be regarded as a consequence of the fluctuating vacuum electromagnetic field \cite{Milonni, Milonni_Aspect_2000}. For the distance between the interacting objects considered in this paper, the relativistic retardation can be ignored. \\

Regarding the sign of the dispersion forces, they can be positive or negative depending on the situation and the geometry. Form the de Boer-Hamaker theorem \cite{Derjaguin_1987,Mahanty}, it follows that the van der Waals forces between two particles of fixed shape and of the same material embedded in a fluid is always attractive. If particles have different composition, the resulting force can be repulsion. This is because in the interaction between the two bodies, an interaction of fluid bodies of same size has to be included. \\

Considering all this, an expression and the sign for the dispersion component of the disjoining pressure $\Pi_\mathrm{w}$ in the context of electroporation, is further presented. This expression is given for a solute particles and a substrate, both in medium. The assumption is that the fluctuating electric field in a dilute solution or gas, acts on each dilute species without distortion of the field by other solute or gas molecules. For such media the Lorentz-Lorenz corrections are unimportant, the Lifshitz formula can be simplified, and $\epsilon$ of the dilute-solution (dilute-gas) limit has the dielectric response strictly proportional to the number of solute (gas) molecules \cite{Derjaguin_1989, Milonni, Parsegian_2006}. \\

The interaction energy of one particle ($\epsilon_\mathrm{s}$) of radius $a$, at a distance $d$ away from a substrate ($\epsilon_\mathrm{hc}$) over a medium ($\epsilon_\mathrm{w}$) is given as \cite{Parsegian_2006} 

\begin{eqnarray}
W \left( {d} \right) = - \frac{kT}{2d^3} {\sum_{n=0}^{\infty}}^\prime \frac{\beta}{4 \pi \epsilon_\mathrm{w}} \left(\frac{\epsilon_\mathrm{hc} - \epsilon_\mathrm{w}}{\epsilon_\mathrm{hc}+ \epsilon_\mathrm{w}} \right)
\label{eq:W_w},
\end{eqnarray}

where ${\beta}/{4\pi \epsilon_\mathrm{w}} = a^3 (\epsilon_\mathrm{s}-\epsilon_\mathrm{w})/(\epsilon_\mathrm{s}-\epsilon_\mathrm{w})$ and all three dielectric permittivities $\epsilon(i\xi)$ are dependent and summed over all of the imaginary frequencies $i\xi_\mathrm{n}$. The prime over the summation indicates that the zero frequency term ($n=0$) is multiplied by $1/2$. The force per unit area, exerted by each ion on the substrate would be a derivative of equation (\ref{eq:W_w}). Then the dispersion component of the disjoining pressure for the number of particles in a segment $\Delta x \approx a$ of unit area, parallel to the substrate, given in terms of the particle's concentration $n(d)$ is 

\begin{eqnarray}
\Pi _\mathrm{w} \left( {d} \right) = -n \frac{ \Delta x a^3}{{d^4}}\frac{3\hbar}
{2}\int\limits_0^\infty {\frac{{\left( {\epsilon _\mathrm{s} - \epsilon_\mathrm{w} } \right) \left( {\epsilon _\mathrm {hc} - \epsilon_\mathrm{w} } \right)}}{{\left( {\epsilon _\mathrm{s} + 2\epsilon_\mathrm{w} } \right) \left( {\epsilon _\mathrm{hc} + \epsilon_\mathrm{w} } \right)}}} d\xi 
\label{eq:Pw_w}.
\end{eqnarray}

For the dispersion part of equation (\ref{eq:W_w}), due to the fact that the electronic polarizabilities are dominant, the typical absorption frequencies $\xi$ are close to the UV region. Then the pre factor $kT$ and the sum over discrete frequencies in equation (\ref{eq:W_w}), are replaced by $\hbar$ and an integral over them \cite{Israelachvili}, thus revealing the quantum nature of equation (\ref{eq:Pw_w}) and the role played by the zero-point energy. Calculating the integral in equation (\ref{eq:Pw_w}), requires the functions $\epsilon(i\xi)$, to be known. Unfortunately, complete data of the absorption spectra of most materials are not available. An estimate of the integral is given in the Appendix.

From equation (\ref{eq:Pw_w}) it can be seen that the sign of the pressure $\Pi_\mathrm{w}$ is determined by the dielectric permittivities for given frequency. The particles will be attracted towards the substrate if $\epsilon_\mathrm{s} > \epsilon_\mathrm{w}$ and $\epsilon_\mathrm{hc} > \epsilon_\mathrm{w}$.

\section{Analytical and numerical estimates of the disjoining pressures $\Delta \Pi$}

\subsection{\label{sec:level3} Analytic estimates of the equilibrium potential}
 
From equation (\ref{eq:ConstRo}), noting that at the surface $E_\mathrm{s}  = \sigma /\epsilon \epsilon _\mathrm{0} $ and in the bulk $\rho _\mathrm{+}   = \rho _\mathrm{-}  $, we get Grahame's equation that relates the known surface charge density $\sigma $ with the unknown surface potential $\psi _\mathrm{s}$ in terms of inverse hyperbolic functions.

\begin{eqnarray}
\psi _s  = \frac{{2kT}}{e}ash\left( {\frac{\sigma }{{\sqrt {8\epsilon \epsilon _0 kT\rho _0 } }}} \right)
\label{eq:Psi_s}.
\end{eqnarray}

From it and from equation (\ref{eq:eqbl}), the surface charge density $\rho_\mathrm{s}$ can be calculated. In the case of thin films, the mid-plane (where $E=0$) is very useful, but this is not applicable in the case of a single membrane.

The potential at any distance $x$ from the surface in Gouy-Chapman Theory 
(high potential) \cite{Israelachvili} is given as

\begin{eqnarray}
\psi \left ( x \right ) \approx \frac{{4kT}}{e} \gamma e^{-\kappa x} 
\label{eq:Psi_x},
\end{eqnarray}

where $\gamma = th\left( {e\psi _\mathrm{s}}/{4kT} \right)$. The Debye length $1/\kappa$ appears as the characteristic decay length of the potential, same as in Debye-H\"{u}ckel theory.

\subsubsection{\label{sec:level2} Numerical simulation}

A coupled simulation of equations (\ref{eq:SE}) and (\ref{eq:PB}) over time of $1$ $\mu$s and simulation space of $0.25\mu$m was done. The bulk ion concentration was considered to be 100mM which corresponds to Debye length of $\approx 1$nm. Microscopic details on the membrane were not considered. It was taken to be a $5$nm thick dielectric slab with dielectric constant of $\epsilon=2$. On both sides of the membrane a Stern layer of thickness $1.8$nm, approximately the size of a water molecule \cite{Israelachvili}, not accessible to diffusing ions, was considered. The dielectric constant in the Stern layer was taken to be the same as the one in the water, $\epsilon=80$. The analytical expressions (\ref{eq:Psi_s}) and (\ref{eq:Psi_x}) were used in the simulation of equations (\ref{eq:SE}) and (\ref{eq:PB})  to find $\psi \left ( x \right ) $ and from it, the initial ion distribution $\rho \left ( x \right ) $ near the membrane. This initial ion distribution was used later in the case when an external electric field was applied. 

The simulation gives a nice agreement and comparable results to Gouy-Chapman theory (high potentials), and the Debye-H\"{u}ckel theory (low potentials). Under an external electric filed there is an ion concentration stabilization near the membrane within the membrane charging time of $\approx 1 \mu$s. This also corresponds to the standard membrane specific capacitance of $c_\mathrm{m} = 1\mu \mathrm{Fcm}^{-2}$. The analytical estimate of the ion concentration change based on the specific capacitance equation (\ref{eq:D_n}) gives a value of $\Delta n \approx 14.4$mM for each increase in the applied voltage of $\Delta \psi_\mathrm{m}=50$mV. For the run time of the simulation ($1 \mu$s) the ion concentration change was $\Delta n \approx 11$mM. 

The type of ions, the bulk ion concentration, and the applied voltages used in the simulation were chosen to resemble the experimental situation of \cite{Melikov_2001}. Besides the agreement, the data from the simulation was used just to illustrate the voltage profile near the membrane in figure (\ref{fig:V}). In all other estimates, standard analytical values were used.

\subsection{\label{sec:level2} Estimate of the gradient of the disjoining pressures}

\subsubsection{\label{sec:level3}The gradient of the electrostatic component of disjoining pressure $\Delta \Pi_\mathrm{e}$}

The electrostatic component of this pressure gradient over the membrane can be calculated using equation (\ref{eq:Pel_apx}), where $\psi_\mathrm{m}$ is the value of the potential far away from the membrane. The applied electric field manifests itself by establishing $\Delta\psi_\mathrm{m}$ across the membrane without substantially affecting the side potentials \cite{Waltz} near the  membrane. The side potentials $\psi_\mathrm{s}$ depend heavily on membranes surface charge and $\psi_\mathrm{s2}=\psi_\mathrm{s1} + \Delta\psi_\mathrm{m}$ as shown in figure (\ref{fig:V}). Then relative to the left side potential, the difference in the electrostatic component of the disjoining pressure on both sides of the membrane is 

\begin{eqnarray}
\Delta \Pi_\mathrm{e} \left( {\psi_\mathrm{m} } \right) = \Pi_\mathrm{left} - \Pi_\mathrm{right} = n_\mathrm{0} \frac{{\left( {ze} \right)^2}}{{kT}}\Delta \psi_\mathrm{m}^2 
\label{eq:DelPel_apx},
\end{eqnarray}

where $\Delta \psi_\mathrm{m}$ is the transmembrane voltage. This expression has not been used yet in the electroporation models and theory. Besides its voltage dependence, away from the membrane, it also reveals its bulk ion concentration dependence that is often missing in the electroporation models. 

In electroporation experiments \cite{Melikov_2001}, it has been noticed that relatively small transmembrane voltage of few hundred mV cause electroporation, in contrast to membrane elasticity model estimates of few volts. The expression (\ref{eq:DelPel_apx}) indicates that $\Delta \psi_\mathrm{m} \approx 300$ mV gives $\Delta \Pi_\mathrm{e}$ over the membrane instability threshold of $\approx 400$ atm. The calculated values of $\Delta \Pi_\mathrm{e}$ are shown in figure (\ref{fig:P}). 


\subsubsection{\label{sec:level3}The gradient of the dispersion component of disjoining pressure $\Delta \Pi_\mathrm{w}$}

Similar to $\Delta \Pi_\mathrm{e}$, due to the externally applied field, the asymmetry in the number of ions on both sides of the membrane ($n_\mathrm{left} = n_\mathrm{eq} + \Delta n$ and $n_\mathrm{right} = n_\mathrm{eq} - \Delta n$), will also cause a pressure gradient in the dispersion component $\Pi_\mathrm {w}$. As indicated from equation (\ref{eq:Pw_w}) van der Waals dispersion force is proportional to the hydrated ion concentration $n(d)$, in a segment $\Delta x$ of unit area, parallel to the membrane. The dispersion force between the ions and the membrane hydrocarbons will give rise to pressure gradient $\Delta \Pi_\mathrm{w}$ due to the difference in the number of ions on both sides of the membrane. 

\begin{eqnarray}
\Delta \Pi_\mathrm{w} = 2\Delta n \left (a^3 \frac{\Delta x}{d^4} \right) I
\label{eq:DelPw_apx},
\end{eqnarray}

where $I$ (in units of energy) is the integral over the imaginary frequencies in equation (\ref{eq:Pw_w}). An estimate of $I$ is given in the Appendix. Further, $\Delta \Pi_\mathrm{w}$ can be estimated for different types of ions. 

$\Delta n$ is proportional with the accumulated charge and an estimate of $\Delta n$ that goes into equation (\ref{eq:DelPw_apx}) can be made through membrane's specific capacitance $c_\mathrm{m}$,

\begin{eqnarray}
\Delta n = \frac{c_\mathrm{m} \Delta \psi_\mathrm{m} }{2 e dx } (1-e^{-t/\tau_\mathrm{ch}})
\label{eq:D_n},
\end{eqnarray}

and for times longer than membrane's charging time $t \gg \tau_\mathrm{ch}$ all of the $\Delta n$ accumulated ions will be 
next to the membrane at a distance of the Stern layer from it. The above expression gives $\Delta n \approx 14.4$mM $\approx 0.2\%$ on one side, for every $\Delta \psi_m=50$mV. When the $\Delta n$ ions are being distributed next to the Stern layer ($d = 1.8 \mathrm{\AA}$), which is about the size of the water molecule diameter \cite{Israelachvili}, $\Delta x$ is in the range of values 

\begin{eqnarray}
d \leq \Delta x \leq d+a
\label{eq:Dx}.
\end{eqnarray}

\begin{figure*}
\includegraphics{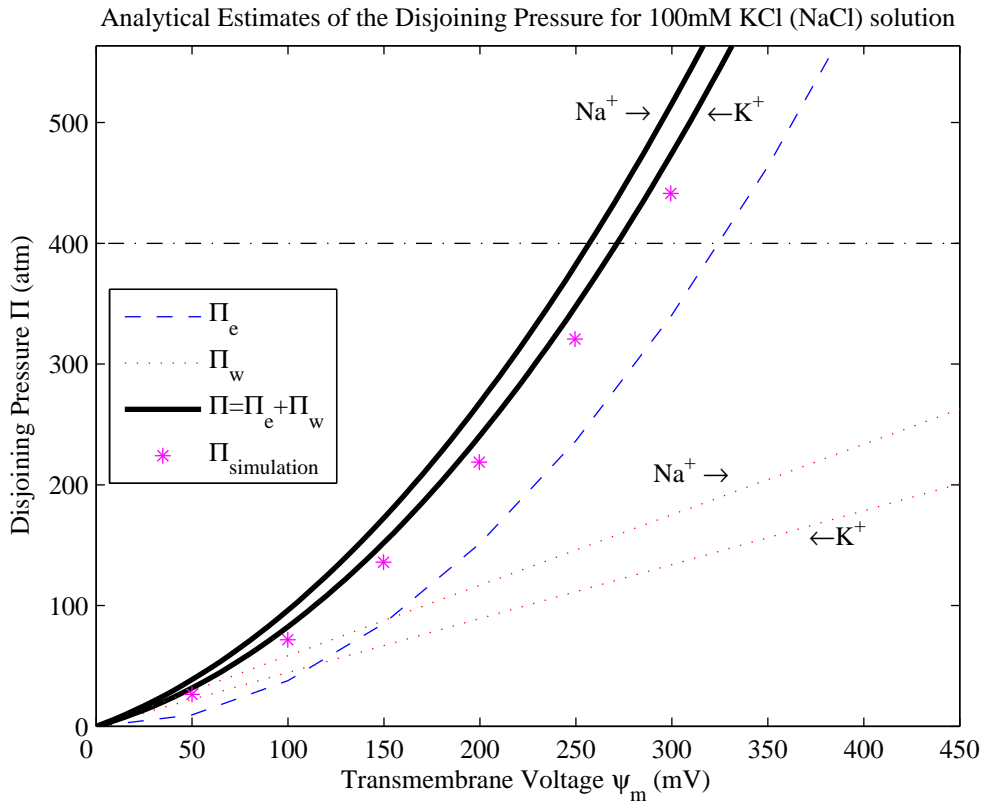}
\caption{\label{fig:wide} Analytical estimates of the electrostatic and dispersion components of the disjoining pressure. Equation (\ref{eq:DelPel_apx}) and (\ref{eq:DelPw_apx}) were used. The increase of the number of ions with the transmembrane voltage is at a rate of $14.4$ mM per $50$ mV ($\Delta n = 14.4 ({\Delta \psi_\mathrm{m}}/{50\mathrm{mV}}) \mathrm{mM} $). The value of $\Pi \approx 400$ atm is indicated as an approximate threshold for membrane instability. If instead of $KCl$ a $NaCl$ salt is used, $\Pi$ will be higher as indicated. Note that $\Pi_\mathrm{w} > \Pi_\mathrm{e}$ for lower values of $\Delta \psi_\mathrm{m}$.}
\label{fig:P}
\end{figure*}

\begin{figure*}
\includegraphics{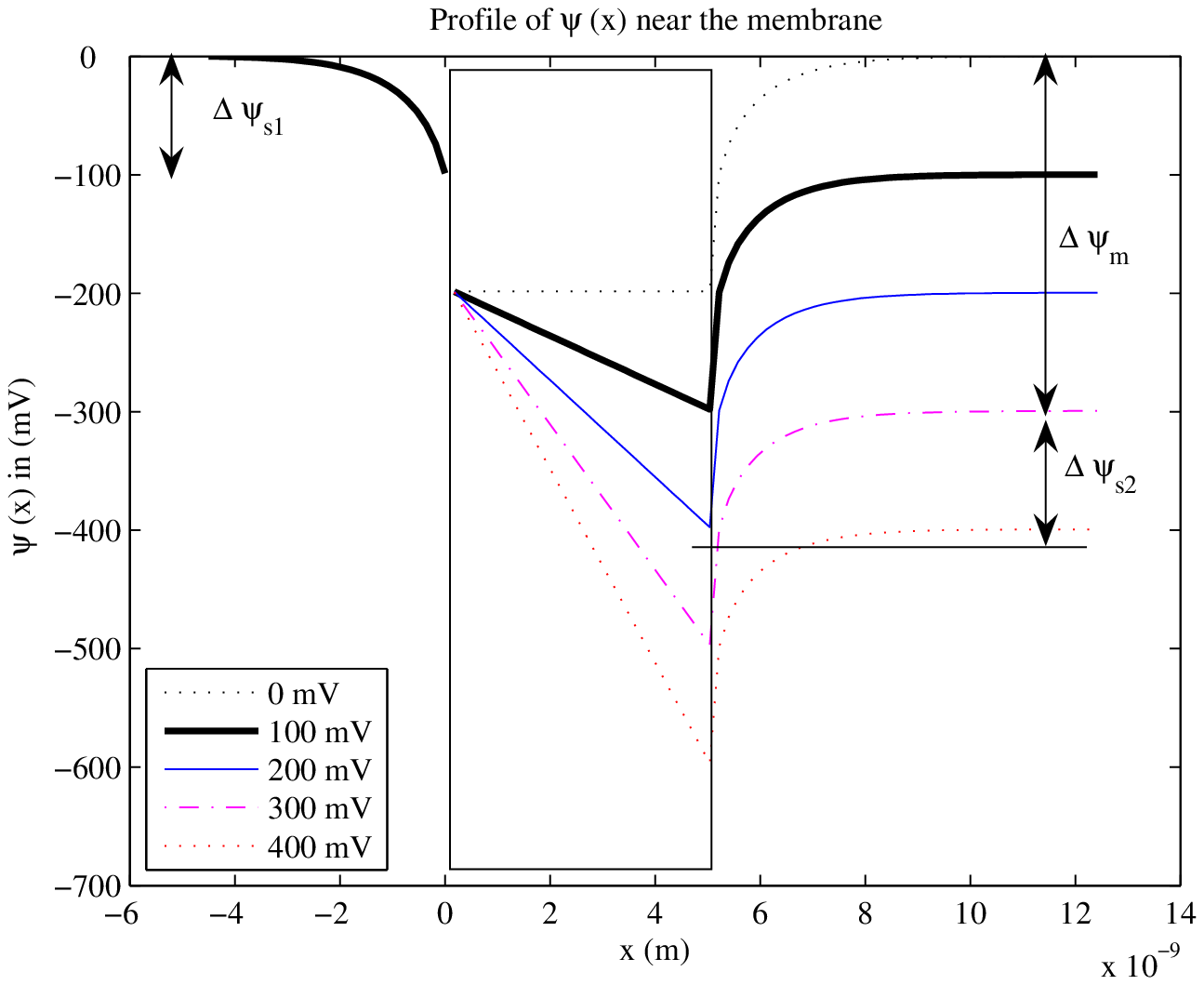}
\caption{\label{fig:wide} Profile of potential $\psi(x)$ across a membrane 
for different applied fields causing transmembrane voltage $\Delta \psi_\mathrm{m}$ of $100, 200, 300$ (indicated) and $400$ mV. On both sides near the membrane, there is an additional jump of $\psi(x) \approx 100$ mV in the Stern layer \cite{Israelachvili}. The Stern potential is not plotted on the left to indicate that $\Delta \psi_\mathrm{m}$ (and the applied field) have virtually no effect on the side potentials $\psi_\mathrm{s}$ \cite{Waltz}. Then it can be noted that $\psi_\mathrm{s2}=\psi_\mathrm{s1}+\Delta \psi_\mathrm{m}$. This fact is used in calculating $\Delta \Pi_\mathrm{e}$ using equation (\ref{eq:DelPel_apx}). This plot was obtained by simulating equations 
(\ref{eq:SE}) and (\ref{eq:PB}) over time of $1$ $\mu$s. Microscopic details in the membrane were ignored and it was considered to be a dielectric slab with surface charge density of $\sigma=0.2$ $\mathrm{Cm}^{-2}$.}
\label{fig:V}
\end{figure*}

According to \cite{Israelachvili}, the hydrated ion radii are $a_\mathrm{Na^{+}}=3.6 \mathrm{\AA}$, and $a_\mathrm{K^{+}}=3.3 \mathrm{\AA}$. Finally, taking the lower limit of equation (\ref{eq:Dx}), for K$^+$ ion concentrations of $\Delta n = 5\times14.4$ mM, corresponding to $\psi_\mathrm{m}=250$ mV we will have a value for the lower limit of $\Delta \Pi_\mathrm{w} = 113$ atm (146 atm for Na$^+$ ions). The values for $\Pi_\mathrm{w}$ (for $\psi_\mathrm{m}=250$ mV) are comparable to $\Pi_\mathrm{e}$, and their sum is in the range of the membrane instability threshold of $\approx 400$ atm. Estimates for wider range of transmembrane voltages are shown in figure (\ref{fig:P}). 

From equation (\ref{eq:DelPw_apx}) it can be seen that the contribution to $\Pi_\mathrm{w}$ of ions at larger distances diminishes rapidly.


\begin{figure*}
\includegraphics{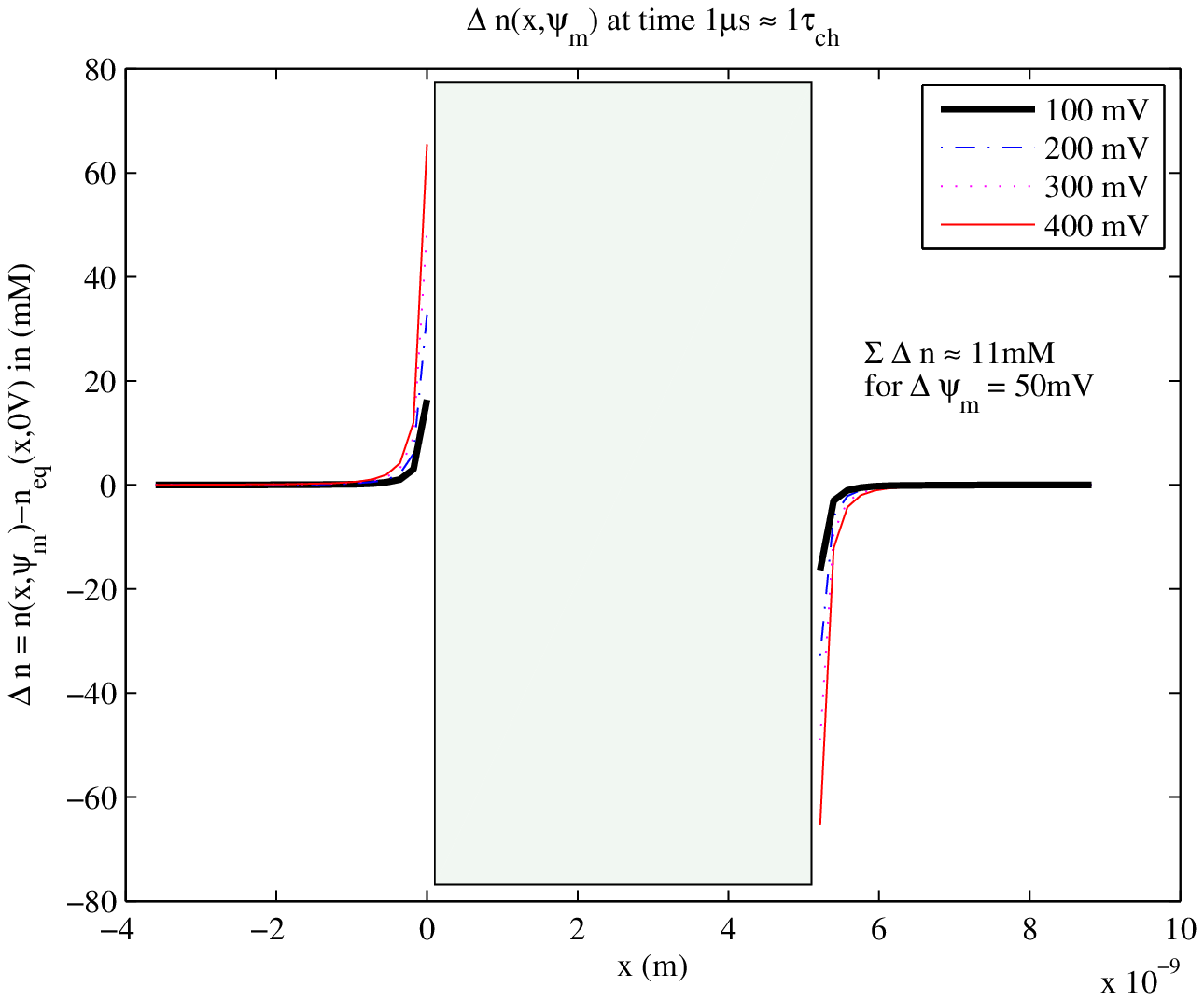}
\caption{\label{fig:wide} Applied electric fields result in ion concentration change of 
$\Delta n(x)$ against the membrane, causing the indicated transmembrane voltages $\Delta \psi_\mathrm{m}$. With time more and more ions are coming to the closest possible approach to the membrane, which is the Stern layer ($\approx 1.8 \mathrm{\AA}$) \cite{Israelachvili}. The simulation was run for time of $1\mu$s and it indicates an ion concentration change of $\approx 11$ mM in the volume left (right) of the membrane for each $50$ mV increase in the transmembrane voltage. An analytical estimate for much longer times, using equation (\ref{eq:D_n}) indicates an accumulation of $\Delta n = 14.4$ mM in the immediate neighborhood of the membrane. These values are used to calculate $\Delta \Pi_\mathrm{w}$, using equation (\ref{eq:DelPw_apx}) and they are shown in figure (\ref{fig:P}).}
\label{fig:Dn}
\end{figure*}

\subsection{\label{sec:level2}General Electroporation Conclusions}

Based on the DLVO approach, the main electroporation aspects now can be interpreted, and few conclusions can be drawn. 

It is been experimentally noticed \cite{Melikov_2001} that relatively small values ($\approx 250$mV) of the transmembrane voltage cause electroporation, which has not been satisfactory explained yet. Arguments using standard DLVO theory (equations (\ref{eq:DelPel_apx}) and (\ref{eq:DelPw_apx})) were presented here, that indicate how a relatively low transmembrane voltage results in large disjoining pressures (hundreds of atmospheres) on the membrane eventually causing electroporation. 

Based on membrane's electrical conductivity during electroporation, experimental evidence strongly suggests the existence of nanometer size defects. This makes electroporation a local effect that involves rearrangement of only few lipid headgroups. The timescale for creation of these defects is of the order of nanoseconds. The only lipid membrane fluctuations of this spatial and temporal scale are the protrusions. This suggests that the disjoining pressure is affecting the protrusion of the lipid headgroups, and is working against their lateral pressure profile. This also suggests that the external electric field compromising the membrane's stability can be localized too.

It takes a short time to create a pore. In contrast it takes a long time for pore destruction (pore lifetime is $\approx ms$ \cite{Melikov_2001}). DLVO gives an insight why this might be the case. Once a nanometer size pore is filed with water, it resembles more the situation of thin films with overlapping interfacial regions. Van der Waals forces reverse sign for this situation and prevent the closing of the pore on nanosecond timescale due to double layer and hydration repulsion, which in the case of thin films, dominate phospholipid membranes as they approach within $0.1$ to $0.3$nm. The interfacial regions of these double layers would depend on the geometrical shape of the pore, which at present is not known and is usually modeled as cylindrical or toroidal. Ion concentration inside the pore will also be different from the bulk concentration. 

The DLVO expressions of the disjoining pressures are bulk concentration and ion type dependent. Expression (\ref{eq:Pel_apx}) has not been used yet in the electroporation models and theory. Besides its voltage dependence, it also reveals its bulk ion concentration dependence that is often missing in the electroporation models.

The cell contains mostly $K^\mathrm{+}$ while the extracellular space contains mostly $Na^\mathrm{+}$ ions. If different ions are present on each side of the membrane, as in the example of a cell, the electroporation models predict that both the pore density and the transmembrane potential will be symmetric. But asymmetrical permeabilization of cell membranes has been experimentally observed \cite{ Teissie_1997, DeBruinKrassowska_1999}. In light of DLVO, the difference in $\Delta \Pi_\mathrm{tot}$ may account for this asymmetry. The disjoining pressures on both sides of the cell will be different, since $\Delta \Pi_\mathrm{w}$ will be different for $K^\mathrm{+}$ and $Na^\mathrm{+}$ counterions. The fact that the cell side facing the anode is being most permeable \cite{ Teissie_1997} is in agreement with the $\Pi_\mathrm{w}$ being larger for $Na^\mathrm{+}$ than for $K^\mathrm{+}$ counterions as seen in figure (\ref{fig:P}). 

Pores live long enough (ms) for lateral diffusion ($\approx 10^{-8}$ m in ms) to cause redistribution of lipids from one side to the other.

Applying DLVO to present electroporation simulation models suggests slight modifications that need to be made. The phenomenological voltage dependent pore creation rate with the constant $B$ that was crudely estimated from Melinkov's experiments \cite{Vasilkoski_2006}, needs to be related to the prefactor in equation (\ref{eq:DelPel_apx}) and equations (\ref{eq:DelPw_apx}) appropriately scaled by a sutable volume. This will make the pore creation rate dependent on the bulk concentration and other material properties of the system contained in the constants, which so far was not the case. This will also slightly change figure (2) in \cite{Vasilkoski_2006} for the lower transmembrane voltage values, and relate the lag time with $\Delta \Pi_\mathrm{w}$ and the alignment of ions against the membrane. Since for $\Pi_\mathrm{w}$ to reach its maximal value, $\Delta n$ really needs to arrive within the Stern layer, next to the membrane, while $\Delta \psi_\mathrm{m}$ and $\Delta \Pi_\mathrm{e}$ do not change substantially. 

Some of these conclusions should be elaborated in more details in a future work.

\section{Conclusions}

The standard DLVO theory, that has been used to describe the stability of thin films, was applied to stability of lipid membranes affected by electric fields. Electroporation is caused by relatively small number of ions ($\approx 1-2\%$), compared to the number of ions already present near the membrane. When external electric field is applied, it breaks the symmetry of the ion distribution in the double layer on both sides of the membrane. This gives a rise to a pressure difference $\Delta \Pi$ in the disjoining pressures on both sides on the membrane. The pressure gradient is causing protrusion of few lipid headgroup compromising the stability of the bilayer. This idea was supported, by estimating the changes in the disjoining pressure $\Delta \Pi_\mathrm{tot}$, and comparing it with lower limit estimates of the bilayer's lateral pressure calculated from statistical dynamics calculations. In comparison with the existing electroporation models, the pore formation energy per lipid volume used in these models corresponds to the Derjaguin's disjoining pressures. With respect to this, slight corrections to the existing electroporation models were also proposed. 

Finally, the dispersion interactions of the double layer counterions and membrane lipid molecules over the Stern layer were introduced in the context of electroporation and membrane stability. They seem to be dominant in the lower range of transmembrane voltages. They are ion size specific and they complement the electrostatic component of the disjoining pressure. Their effect on metal surfaces, for example, can be quite large, since expression (\ref{eq:Int}) for metals is at least an order of magnitude larger. This can have significant effects on metal solution interfaces like electrodes in batteries, fuel cell membranes, electrochemical dealloying and similar situations. 

Experimental evidence points out that electroporation starts to occur within the range of transmembrane voltages that cells are routinely experiencing. This is just above the range of voltages involved in neural signal transmission (less than $200$mV). Pores are created on a nanosecond timescale, but they persist for milliseconds. Thus, it seems that the undesirable electroporation placed a limit on the range of transmembrane voltages involved in neural signal transmission.

Main limitations of the standard DLVO approach presented here are the facts that Poisson-Boltzmann and Lifshitz theories are (continuous) mean field theories. Also not taken into account are few other effects, as the steric effects (due to finite ion sizes), image forces, solvation (hydration) forces, and discreteness of surface charges, between other effects. This may be a subject of future work. 

\begin{acknowledgments}
I wish to acknowledge the help and the valuable advises of Anatoly Adamov, Robert S. Cantor, Evan Evans, Richard Templer, David Weaver, Peter Tieleman, and Vasilka \v{S}opova.
\end{acknowledgments}

\appendix*

\section{Estimate of $\Delta \Pi_\mathrm{w}$}

To calculate exactly the integral in equation (\ref{eq:Pw_w}), the functions 
$\epsilon(i\xi)$, need to be known. Complete set of data of the absorption spectra of most materials are not available. Reliable results can be obtained by several main bands in the absorption spectra, or to a first approximation, the visible and the near ultraviolet bands \cite{Derjaguin_1987}.

All the data in the following estimate is from reference \cite{Israelachvili}. 
Calculating the integral in equation (\ref{eq:Pw_w}), for the dielectric permittivity in the visible and the ultraviolet range that are dominant in the given situation, the following estimate is obtained. 

\begin{eqnarray}
\epsilon_\mathrm{hc} = 1+ \frac{n_\mathrm{hc}^2-1}{1 + \xi^2 / \xi_\mathrm{e}^2}, 
\\
\label{eq:eps_hc}
\epsilon_\mathrm{w} = 1+ \frac{n_\mathrm{w}^2-1}{1 + \xi^2 / \xi_\mathrm{e}^2}, \\
\label{eq:eps_w}
\epsilon_\mathrm{s} = 1 + \xi^2 / \xi_\mathrm{se}^2
\\
\label{eq:eps_xi},
\end{eqnarray}

where, according to \cite{Israelachvili}, the refractive indices are  $n_\mathrm{w}=1.33$, $n_\mathrm{hc}=1.41$, and the ion's dielectric permittivity $\epsilon_\mathrm{s}$ is routinely approximated as that of a metal. The plasma frequencies of the free electron gas are $\xi_\mathrm{e}=3.0 \times 10^{15} s^{-1}$ and $\xi_\mathrm{se}=5.0 \times 10^{15} s^{-1}$. Using \ref{eq:eps_hc}, \ref {eq:eps_w}, and \ref {eq:eps_xi} in equation (\ref{eq:Pw_w}), and solving the integral for $\xi$ gives the following value

\begin{eqnarray}
I = \frac{3\hbar}{2}\int\limits_0^\infty {\frac{{\left( {\epsilon _\mathrm{s}  - \epsilon_\mathrm{w} } \right) \left( {\epsilon _\mathrm {hc} - \epsilon _\mathrm{w} } \right)}}{{\left( {\epsilon _\mathrm{s} + 2\epsilon _\mathrm{w} } \right) \left( {\epsilon _\mathrm{hc} + \epsilon _\mathrm{w} } \right)}}} d\xi = 2.1 \times 10^{-20} \mathrm{J}
\label{eq:Int}.
\end{eqnarray}

This value of $I$ is further used to estimate $\Delta \Pi_\mathrm{w}$ for specific ion types and concentrations. It should be noted that this is just an order of magnitude estimate, considering the imprecision of the quantities going into this estimate, and the mean field approach.
~~~~~~~~~~~~~~~~~~~~~~~~~~~~

\newpage 

\bibliography{EPT}

\end{document}